\documentstyle[12pt]{article}
\input epsf
\newcommand{\NPB}[3]{\emph{ Nucl.~Phys.} \textbf{B#1} (19#2) #3}   
\newcommand{\PLB}[3]{\emph{ Phys.~Lett.} \textbf{B#1} (19#2) #3}   
\newcommand{\PRD}[3]{\emph{ Phys.~Rev.} \textbf{D#1} (19#2) #3}

\def\dalemb#1#2{{\vbox{\hrule height .#2pt
        \hbox{\vrule width.#2pt height#1pt \kern#1pt
                \vrule width.#2pt}
        \hrule height.#2pt}}}

 \def\bd{\begin{document}} \def\ed{\end{document}}
\def\ds{\documentstyle} \let\fr=\frac \let\bl=\bigl \let\br=\bigr
\let\Br=\Bigr \let\Bl=\Bigl 
\let\bm=\bibitem
\let\na=\nabla
\let\pa=\partial \let\ov=\overline
\def\ie{{\it i.e.\ }} 
\newcommand{\pr}{\paragraph{}}
\newcommand{\be}{\begin{equation}}
\newcommand{\ee}{\end{equation}}
\newcommand{\beba}{\begin{equation}\begin{array}{lcl}}
\newcommand{\eaee}{\end{array}\end{equation}}
\newcommand{\bea}{\begin{eqnarray}}
\newcommand{\eea}{\end{eqnarray}}
\newcommand{\ba}{\begin{array}}
\newcommand{\ea}{\end{array}}
\newcommand{\td}{\tilde}
\newcommand{\norsl}{\normalsize\sl}
\newcommand{\ns}{\normalsize}
\newcommand{\refs}[1]{(\ref{#1})}
\def\simlt{\mathrel{\lower2.5pt\vbox{\lineskip=0pt\baselineskip=0pt
           \hbox{$<$}\hbox{$\sim$}}}}
\def\simgt{\mathrel{\lower2.5pt\vbox{\lineskip=0pt\baselineskip=0pt
           \hbox{$>$}\hbox{$\sim$}}}}

\begin{document}
\thispagestyle{empty}
\rightline{\large\sf hep-ph/9905311}
\rightline{CPHT-S717.0599}
\rightline{CERN-TH/99-128}
\rightline{IEM-FT-192/99}
\rightline{\large May 1999}
\vskip 1.0truecm
\centerline{\Large\bf Direct collider signatures of large extra 
dimensions~\normalsize\footnote{Work partly supported by the EU under 
TMR contract ERBFMRX-CT96-0090 and by CICYT of Spain under contract 
AEN98-0816.}}
\vskip 1.truecm
\centerline{{\large\bf I. Antoniadis}~$^\dag$, 
{\large\bf K. Benakli~$^\ddag$} and {\large\bf M. Quir\'os}~$^\natural$}
\vskip .5truecm
\centerline{{\it $^\dag$Centre de Physique Th{\'e}orique}
\footnote{Unit\'e mixte du CNRS UMR 7644.}}
\centerline{\it Ecole Polytechnique, 91128 Palaiseau, France}
\vskip .5truecm
\centerline{{\it $^\ddag$CERN Theory Division
 CH-1211, Gen\`eve 23, Switzerland}}
\vskip .5truecm
\centerline{{\it $^\natural$Instituto de Estructura de la Materia}}
\centerline{\it CSIC, Serrano 123, 28006 Madrid, Spain}

\vskip 1.truecm
\centerline{\bf\small ABSTRACT}
\vskip .5truecm

The realization of low (TeV) scale strings  usually
requires the existence of large (TeV) extra dimensions where 
gauge bosons live. The direct production of Kaluza--Klein excitations of the photon and 
$Z$--boson at present and future colliders is studied in
this work. At the LEPII, NLC and Tevatron colliders, these Kaluza--Klein modes lead to 
deviations from the standard model cross-sections, which provide lower bounds 
on their mass. At the LHC the corresponding resonances can be produced and 
decay
on-shell, triggering a characteristic pattern in the distribution of dilepton 
invariant mass.

\hfill\break
\vfill\eject

\section*{Introduction}

There has been recently a lot of interest in the possibility that string 
theories become relevant at low energies, accessible to future 
accelerators \cite{KK-TeV}--\cite{KK-TeV3}. This is realized for instance if 
the string tension is in the TeV range~\cite{TeV}, which is also motivated 
by an alternative solution to the gauge hierarchy 
problem~\cite{hierar,hierar2}. 
The realization of this idea in weakly coupled type~I theories~\cite{typeI} 
implies the existence of large extra dimensions,
in the millimetre to Fermi range. These dimensions are seen only by
gravity, while the standard model interactions are confined to (D+3)-branes 
transverse to them. One of the main predictions of these theories is that 
gravity becomes strong in the TeV region, and so it can have important 
observable effects in particle accelerators at energies close to the string 
scale~\cite{lowgrav}. 

Besides this effect from the transverse dimensions, there may be some 
longitudinal dimensions (seen by gauge interactions), which are also in the 
TeV range~\cite{KK-TeV}--\cite{Mar}. 
Such dimensions are always present, with the only 
exception of having six transverse dimensions at the Fermi scale. 
Moreover, they can have a size a bit larger than the string length, 
which is motivated either 
by supersymmetry breaking through the 
compactification [1--4, 10, 11], or by power-law 
running of 
couplings~\cite{power-law} to achieve unification of gauge 
interactions~\cite{unif2,unif3}, or by anisotropic 
compactifications~\cite{anisotropic}. 

A scenario similar to the type I strings where the longitudinal dimensions are 
present might also be studied in 
Ho\v rava--Witten compactifications of M-theory~\cite{HW,Mth}.
On the other hand, the realization of TeV strings in weakly coupled type~II 
theories does not require the existence of large transverse dimensions 
since, in these theories, the weakness of gravitational interactions
can be accounted for by the smallness of the string coupling~\cite{II}. 
They also allow for a scenario with the string scale at intermediate energies
while keeping two extra-dimensions at the TeV.
As a result, there are no strong quantum gravity effects at TeV energies and 
the only observable effects below the string scale are due to the presence of 
longitudinal dimensions.

In this work we study virtual effects and possible on-shell production of 
Kaluza--Klein (KK) excitations of gauge bosons in present and future accelerators. 
Production of a single KK excitation assumes
non-conservation of the momenta along the extra--dimension. 
This possibility is realized in a large class of 
models, where quarks and leptons are localized at 
particular points of the internal (longitudinal) space, similar 
to the twisted states of the heterotic string. We restrict our analysis
to such models and leave for a future study those where quarks and 
leptons  also have  KK-excitations.

For the case of one large extra dimension we find 
that existing colliders, LEPII and Tevatron, will be able to exclude
compactification scales less than  $\sim$ 1.9 TeV. Low--energy precision 
data~\cite{NY,MP,Mar} seem to provide stronger constraints 
$\sim$ 2.5 TeV. This is because of the large number of events that allow us 
to have a very high precision on the width of $Z$ boson and the Fermi constant.
These results imply that it is unlikely that we observe effects of KK excitations 
of gauge bosons at present colliders.
 
The Next Linear Colliders (NLC) with centre--of--mass energy of 500 GeV and 
1 TeV will easily probe (through virtual effects) 
compactification scales up to around 8 TeV and 13 TeV, respectively. 
At the LHC we
find that  the non-observation of 
deviations from the standard model prediction of the total number of 
lepton pairs
with centre--of--mass energy above 400 GeV would translate into a lower bound 
of the order of 6.7 TeV and 9 TeV, 
for one and two large extra--dimensions, respectively. Hadronic colliders allow 
the  most exciting possibility of direct production of KK excitations by 
Drell--Yan processes~\cite{KK-TeV3}. A characteristic signature will 
be the appearance of two overlapping narrow resonances, corresponding to the 
first excited state of the photon and the $Z$ gauge boson, for every 
longitudinal large dimension.  Our analysis indicates 
that in the case of one dimension (or many dimensions with the same 
values of 
the compactification scales) it is likely that only the first peak 
could be seen in the first run. However 
this scenario can be differentiated from 
$Z'$ models because of the presence of excitations of $W$ bosons. 
The present analysis completes
the original work of the authors~\cite{KK-TeV3} where specific couplings 
were assumed.

\section*{Production at $e^+e^-$ colliders}

The exchange of KK excitations of the photon and $Z$ boson leads to 
modifications
of the cross section $e^+e^-\rightarrow \mu^+\mu^-$. As  the invariant
mass of the produced fermion pairs is (to a first approximation) of the order
of the machine energy, resonances cannot be directly observed unless
the machine energy happens to be very close to the mass of one of
the excitations, or else a scanning of energies is made. Because of the
clean environement, these experiments allow the performance of high--precision tests
and the extraction of bounds on the new physics.   

The total cross section for the annihilation of unpolarized
electron-positron pairs $e^+e^-$, with a centre--of--mass energy $\sqrt s$,
to lepton pairs  $l^+l^-$, through the exchange of vector
bosons in the s-channel, is given by:

\begin{equation}
\sigma_T^0(s)={s\over 12 \pi} \sum_{\alpha ,\beta=\gamma, Z, KK}g^2_{\alpha}
({\sqrt s}) g^2_{\beta} ({\sqrt s}) {(v^{\alpha}_e v^{\beta}_e+
a^{\alpha}_e a^{\beta}_e)(v^{\alpha}_l v^{\beta}_l + a^{\alpha}_l
a^{\beta}_l) \over (s -m^2_{\alpha} + i\Gamma{_\alpha}
m_{\alpha})(s-m^2_{\beta} - i\Gamma_{\beta} m_{\beta})} \ ,
\label{eesig}
\end{equation}
where the labels $\alpha ,\beta$ stand for the different neutral vector
bosons $\gamma$, $Z$, and their KK excitations with coupling constants
$g_{\alpha}$ and masses $m_\alpha$ given by
\begin{equation}
\label{masasKK}
m_n^2=m_0^2+\frac{\vec{n}^2}{R^2}\, .
\end{equation}
Here, $R$ denotes the common radius for $D$ large (TeV) dimensions and 
$m_0$ is a $(D+4)$-dimensional mass which, for instance, can appear through an
ordinary Higgs mechanism, such as that responsible for the electroweak
symmetry breaking. The widths $\Gamma_\alpha$ are decay rates into 
standard model fermions $f$:
\begin{equation}
\Gamma\left(X_n\rightarrow f\bar{f}\right)
=g^2_\alpha \frac{m_n}{12\pi}C_f(v_f^2+a_f^2)
\label{Gammaf}
\end{equation}
and their scalar superpartners
\begin{equation}
\label{Gammasf}
\Gamma\left(X_n\rightarrow \widetilde{f}_{(R,L)} 
\widetilde{\bar{f}}_{(R, L)}\right)
=g^2_\alpha\frac{m_n}{48\pi}C_f(v_f\pm a_f)^2\, ,
\end{equation}
where $C_f = 1$ (3) for colour singlets (triplets) and $v_f, a_f$ are the 
standard model vector and axial couplings. The precise value of the width 
is not important in most of our analysis based on virtual effects. 
It is however important in the case of on-shell production of KK excitations. 
In our studies we use the standard model particles as the only accessible 
final states. If instead one includes also their superpartners, then the 
widths of KK excitations of the photon and $Z$ are multiplied by a factor 3/2. As a result the resonances are wider. 

We expect, in experiments to be held at energies far below the resonance 
peaks, that interference terms will be the main source of any observable 
effects. As the latter are small, they must be computed with the highest 
possible accuracy. In our analysis we include radiative corrections, and
in particular the bremsstrahlung effects on the initial electron and
positron~\cite{Dlr}. These are described by the convolution of (\ref{eesig}) 
with radiator functions, which describe the probability of having a
fractional energy loss, $x$, due to the initial--state radiation:
\begin{equation}
\sigma_T(s) =\int^{x_{\rm max}}_0 dx \sigma_T^0(s') r_T(x) ,\quad s'=s(1-x)
\label{eesigr}
\end{equation}
In the above equation, $x_{\rm max}$ represents an experimental cut off for
the energy of emitted soft photons in bremsstrahlung processes. The
radiator function is given by~\cite{Dlr}:

\begin{equation}
\label{radiator}
r_T(x) =(1 + X) y x^{y -1}+ H_T(x)\ ,
\end{equation}
with:
\bea
X &=& {e^2(\sqrt{s}) \over 4\pi^2} \left[{\pi^2\over 3}- {1\over
2}+{3\over 2}\left(\log{{s\over m_e^2}}-1\right)\right]\nonumber \\ 
y &=&{2e^2(\sqrt{s}) \over 4\pi^2} \left(\log{{s\over m_e^2}}-1\right)
\nonumber \\ 
H_T &=& {e^2(\sqrt{s}) \over 4\pi^2} \left[{1+(1-x)^2\over x}\left(\log{{s\over
m_e^2}}-1\right)\right]-{y\over x}\ ,
\eea
where $m_e$ is the electron mass.

\begin{figure}[htb]
\centering
\epsfxsize=4.5in
\hspace*{0in}
\epsffile{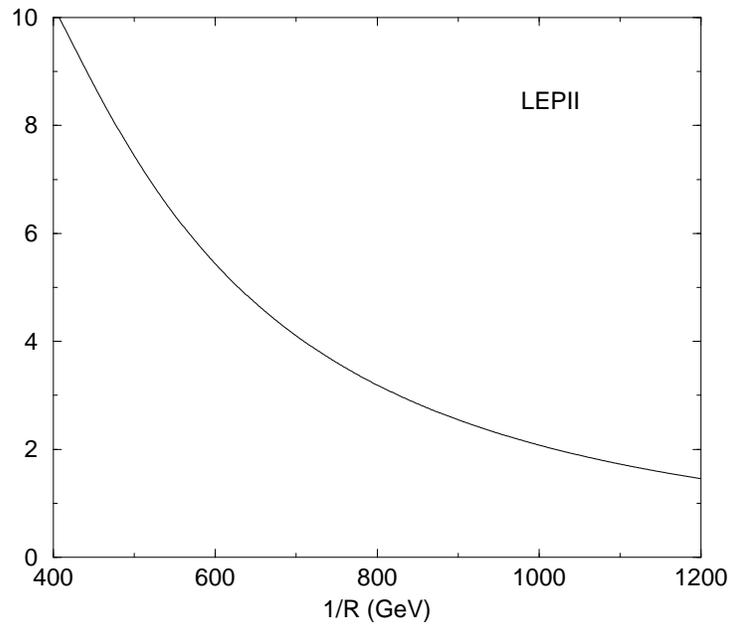}
\caption{\it Ratio $\left|{N_T(s)-N_T^{\rm SM}(s) \over \sqrt{N_T^{\rm SM}(s)}}
\right|$ from the total cross section at LEPII. We assumed a luminosity 
times efficiency of 200 $pb^{-1}$.}
\label{fig:fig1}
\end{figure}

We use ${\sqrt s}=189$ GeV for LEPII, 500 GeV for NLC-500 and 1 TeV for
NLC-1000,  together with the numerical values for the experimental 
cuts:
\begin{eqnarray}
x_{\rm max}({\rm LEPII})&=& 0.77,\nonumber\\
x_{\rm max}({\rm NLC\rm{-}500})&=& 0.967, \nonumber\\
x_{\rm max}({\rm NLC\rm{-}1000})&=&0.992,
\end{eqnarray}
coming from the condition of removing the $Z$--boson tail, which
amounts to imposing the cut $s'\ge M_Z^2$.
\begin{figure}[htb]
\centering
\epsfxsize=4.5in
\hspace*{0in}
\epsffile{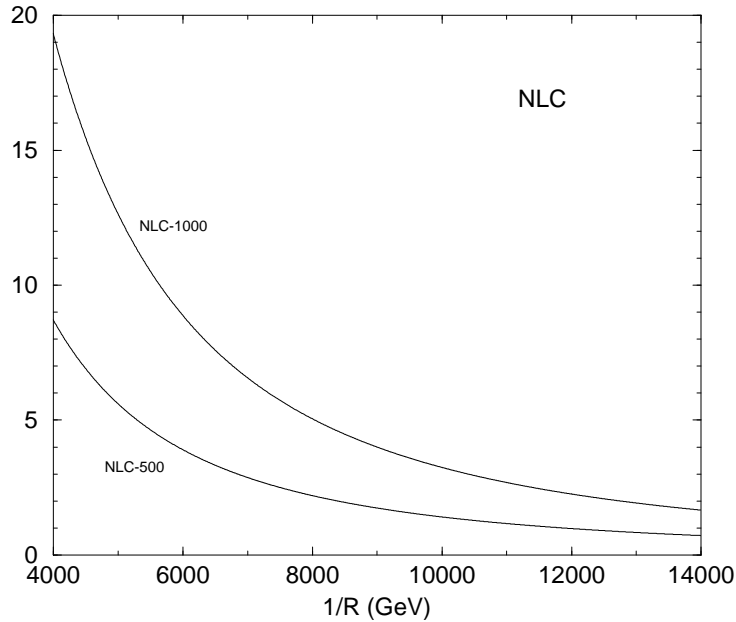}
\caption{\it Ratio $\left|{N_T(s)-N_T^{\rm SM}(s) \over \sqrt{N_T^{\rm SM}(s)}}
\right|$ from the total cross section at NLC-500 and NLC-1000. 
We assumed a luminosity times efficiency of 75 $fb^{-1}$ and 
200 $fb^{-1}$, respectively.}
\label{fig:fig2}
\end{figure}

We first consider the case of one extra dimension ($D=1$). 
The sum over KK modes
in (\ref{eesig}) converges rapidly and it is dominated by the lowest modes.
We then use the approximation where the couplings of all $n>0$ modes are
equal~\footnote{Summing modes with $n$ positive, the coupling of $n\neq 0$
modes is enhanced by a factor $\sqrt{2}$~\cite{SS6}.}. 
In Fig.~\ref{fig:fig1} we plot, for LEPII, the ratio
\begin{equation}
\Delta_{T}=\left|{N_T(s)-N_T^{\rm SM}(s) \over \sqrt{N_T^{\rm SM}(s)}}
\right|
\end{equation}
where $N_T(s)$ is the total number of events while $N_T^{\rm SM}(s)$ is 
the corresponding quantity expected from the standard model. This quantity 
gives an estimate of the deviation from the background fluctuation in the 
case of a large number of events. The corresponding plots for NLC-500 
and NLC-1000 are given in Fig.~\ref{fig:fig2}.

We can see from Fig.~\ref{fig:fig1} that a measurement of the total number of 
events (from which one extracts $\sigma_T(s)$) puts a limit of 1~TeV
(95\% c.l.) from LEP-II~\footnote{Combining data from the four LEP experiments
would lead to a corresponding bound of $\sim$ 1.9 TeV.}. 
Future experiments at the NLC with centre--of--mass energies of 500 GeV and 1 TeV, 
and luminosities of 75 fb$^{-1}$ and 200 fb$^{-1}$, will allow us to probe sizes 
of the order of 8 TeV and 13 TeV, respectively, as can be seen from
Fig.~\ref{fig:fig2}.

\section*{Production at hadron colliders}

At hadron colliders, the KK excitations
might be directly produced in Drell--Yan processes $pp \rightarrow
l^+l^-X$ at the LHC, or $p{\bar p} \rightarrow l^+l^-X$ at the Tevatron, with
$l=e,\mu,\tau$. This is due to the fact that the lepton pairs are
produced via the subprocess $q{\bar q}\rightarrow l^+ l^- X$ of 
centre--of--mass energy $M$. We will follow here the method of Ref.~\cite{Paco} where
a similar analysis was performed for the production of $Z'$ vector bosons
from $E_6$ string-inspired models.

The two colliding partons take a fraction
\begin{equation}
x_a={M \over \sqrt s}\ e^{y} \quad{\rm and}\quad
x_b={M \over \sqrt s}\ e^{-y}
\end{equation}
of the momentum of the initial proton ($a$) and (anti)proton ($b$), with a
probability described by the quark or antiquark distribution functions
$f^{(a)}_{q,\bar q}(x_{a}, M^2)$ and $f^{(b)}_{q,\bar q}(x_{b}, M^2)$.

The total cross section, due to the production is given by:
\begin{equation}
\sigma= \sum_{q={\rm quarks}} \int^{\sqrt s }_0 dM \int^{\ln (\sqrt s
/M)}_{\ln (M/\sqrt s)}dy \ g_q (y, M) S_q (y, M) \ ,
\end{equation}
where
\begin{equation}
g_q (y, M)= {M \over 18\pi} x_a x_b \ [f^{(a)}_q (x_a,M^2)
f^{(b)}_{\bar q} (x_b, M^2) + f^{(a)}_{\bar q} (x_a, M^2) f^{(b)}_q (x_b,
M^2)]\ ,
\end{equation}
and
\begin{equation}
S_q (y, M)= 
\sum_{\alpha ,\beta\gamma, Z, KK}g^2_{\alpha}(M) g^2_{\beta}(M) 
{(v^{\alpha}_e v^{\beta}_e+
a^{\alpha}_e a^{\beta}_e)(v^{\alpha}_l v^{\beta}_l + a^{\alpha}_l
a^{\beta}_l) \over (s -m^2_{\alpha} + i\Gamma{_\alpha}
m_{\alpha})(s-m^2_{\beta} - i\Gamma_{\beta} m_{\beta})} \ .
\end{equation}

In our computations we used leading--order approximations (set 3 of CTEQ parton
distribution functions~\cite{CTEQ}) and we included a multiplicative
$K$-factor~\cite{kf}. We take $K=1.3$ for the Tevatron and $K=1.1$ for the LHC. 

\subsection*{One extra dimension}

\begin{figure}[htb]
\centering
\epsfxsize=4.5in
\hspace*{0in}
\epsffile{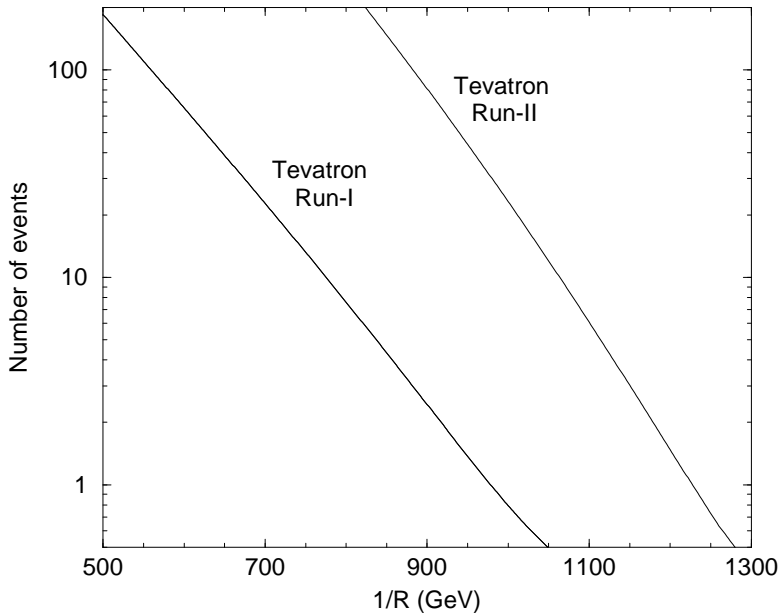}
\caption{\it Number of $l^+ l^-$-pair events with centre--of--mass 
energy above 400 GeV (600 GeV) expected at the Tevatron run-I (run-II) 
with integrated luminosity $\int {\cal L}dt= 110\ pb^{-1}$
($\int {\cal L}dt= 2\ fb^{-1}$) and efficiency times acceptance of
$\sim$ 50\%, as a function of $R^{-1}$.}
\label{fig:fig3}
\end{figure}

Let us first consider, as we did for the case of $e^+ e^-$-colliders, the
simplest case of one extra dimension. At the Tevatron, the CDF collaboration 
has collected an integrated luminosity $\int {\cal L}dt= 110\ \rm{pb}^{-1}$ 
during the 1992-95 running period. From the non-existence of candidate 
events at $e^+e^-$ invariant mass above 400 GeV a lower bound on the 
size of the extra dimension can be derived.
In Fig.~\ref{fig:fig3} we plot the number of expected events 
assuming efficiency (times acceptance) of $\sim$ 50\% in this region. 
This leads to a limit of $R^{-1}\simgt $ 900 GeV with a 95\% confidence 
level. An efficiency of $\sim$ 90\% would have led to $R^{-1}\simgt$ 940 GeV.
%
In Fig.~\ref{fig:fig3} we also plot the expected number of events in the 
run-II of the Tevatron with a centre--of--mass energy $\sqrt{s}=2$ TeV. 
We use an integrated luminosity $\int {\cal L}dt= 2\ fb^{-1}$ with an 
efficiency of $\sim$ 50 \%. The non-observation of any candidates would 
translate to a limit of $R^{-1}\simgt 1.2$ TeV.

\begin{figure}[htb]
\centering
\epsfxsize=4.in
\hspace*{0in}
\epsffile{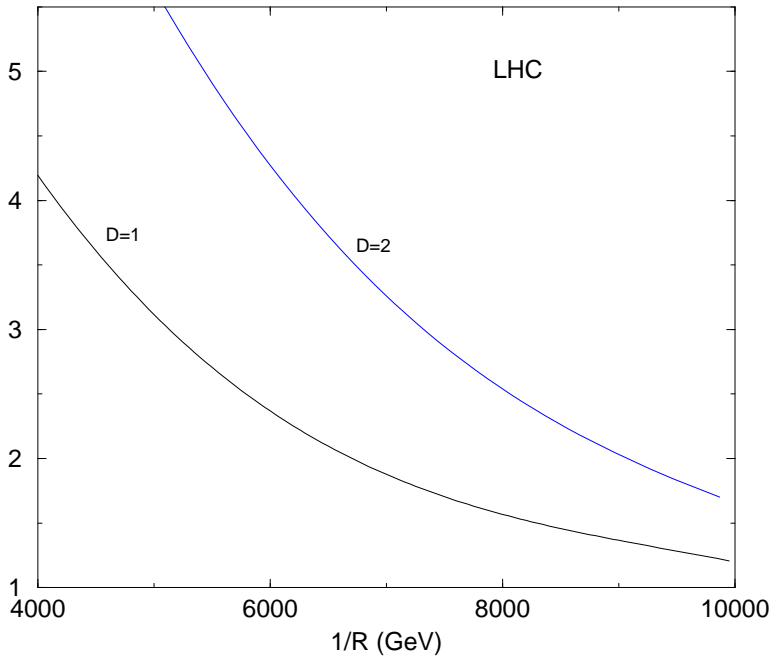}
\caption{\it Ratio 
$\left|{N_T(s)-N_T^{\rm SM}(s) \over \sqrt{N_T^{\rm SM}(s)}}\right|$ from the 
total cross section at the LHC for dilepton--invariant mass above 400 GeV as a 
function of $R^{-1}$. We assumed a luminosity (times acceptance) 
times efficiency of 5 $fb^{-1}$.}
\label{fig:fig4}
\end{figure}

Certainly more promising for probing TeV-scale extra-dimensions
are the LHC future experiments at $\sqrt s =14$ TeV. In Fig.~\ref{fig:fig4}
we plot the deviation from the standard model prediction of the total number of
lepton pairs with center of mass energy above 400 GeV. Non-observation of
such deviations would translate into a lower bound for $R^{-1}$ of 6.7~TeV
(95\% c.l.). One could instead cut off the lepton pairs with 
centre--of--mass energy at a scale $\sim$ 2.5 TeV, above which no event is expected from 
standard model interactions. A  4.8 TeV limit (95\% c.l.) on the scale of 
compactification is obtained from not having at least three candidate events  
above 2.5 TeV.

Within this range of TeV values, the two KK excitations of the photon and $Z$-boson are very close to each other and would be difficult to separate. Moreover
the second, and higher, peaks are very small and likely to be missed by the 
first round of the experiment. In Fig.~\ref{fig:fig5} 
we plot the expected shape of a resonance that could be discovered for an 
extra dimension of 3 TeV.

\subsection*{More than one extra dimensions}

In this section we generalize the previous results concerning the 
production of KK excitations at large hadron colliders to the case of 
more than one large extra dimensions. 

For simplicity, in our numerical analysis, we take the large longitudinal 
dimensions to have a common compactification radius $R$. 
Our computation involves the sum:
\begin{equation}
\sum_{|\vec{n}|} \frac {g^2{(|\vec{n}|)}}{|\vec{n}|^2}
\label{sumdiv}
\end{equation}
coming from interference terms between the exchange of the photon and
$Z$-boson and their KK excitations, where $g^2(|\vec{n}|)$ are the KK-mode 
couplings.

In the case of one extra-dimension the sum in (\ref{sumdiv})  converges 
rapidly. Only the first few terms contribute  and it is legitimate to 
assume that all the KK modes have the same couplings to the boundary fields. 
This is not the case for two or more dimensions as (\ref{sumdiv}) is 
divergent and needs to be regularized. The tree--level dependence of 
$g(|\vec{n}|)$ on $|\vec{n}|$ can be computed in a string framework and 
leads to an exponential drop-off~\cite{{KK-TeV22},{bachas}}:
\begin{equation}
g(|\vec{n}|)\sim g\; a_{(|\vec{n}|)}\; 
e^{\frac {-c|\vec{n}|^2}{2R^2 M_s^2}}\, ,
\end{equation}
where $M_s$ is the string scale, $c$ is a constant and  $a_{(|\vec{n}|)}$ 
takes into account the normalization of the gauge kinetic term, as only 
the even combination couples to the boundary. For the case of two 
extra-dimensions $a_{(0,0)} =1$, $a_{(0,p)} =a_{(q,0)}=\sqrt{2}$ and 
$a_{(q,p)} =2$ 
with $(p,q)$ positive ($>0$) integers. Possible higher--order (loop) 
corrections to KK-mode couplings are assumed to be small.

\begin{figure}[htb]
\centering
\epsfxsize=4.5in
\hspace*{0in}
\epsffile{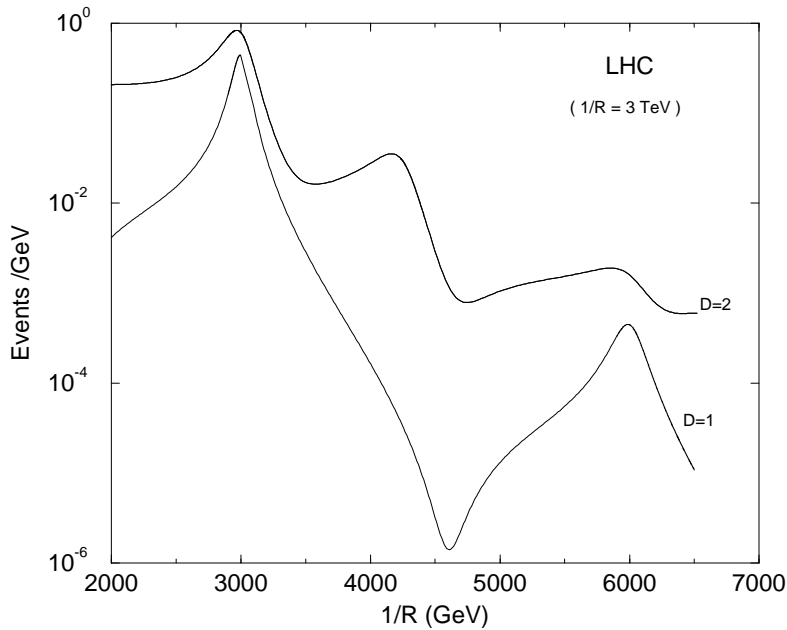}
\caption{\it First resonances in the LHC experiment due to a KK state for one
and two extra dimensions at 3 TeV.}
\label{fig:fig5}
\end{figure}

The behaviour of the cross-sections in the case of more dimensions have
some peculiarities that  we would like to mention. When the number of 
extra dimensions increases, there is an increasing degeneracy of states 
within each mass level that leads to bigger resonances. Moreover, the 
spacing between the mass levels decreases, which means that more resonances 
could be reached by a given hadronic machine. There is an enhancement of the 
effects of KK states, which allows us to probe smaller values of the radius.
Another difference with the one-dimensional case is the sensitivity 
to the value of the string scale as the latter determines where the 
divergent sum is cut off. In the case of non-degenerate radii, a 
corresponding number of resonances due to the first KK excitation along 
each direction can be observed with no regular spacing patterns.

In Fig.~\ref{fig:fig4} we plot the deviation from the standard model 
prediction of the total number of lepton pairs with centre--of--mass energy 
above 400 GeV for the case of two extra dimensions of degenerate radii. 
Non-observation of such deviations would translate into a lower bound 
of 9~TeV (95\% c.l.). In Fig.~\ref{fig:fig5}  we 
plot the expected shape of resonances that could be discovered for two 
extra dimensions of 3~TeV. In our examples we 
arbitrarily choose $M_s \sim 6/R$ and we include only standard model particles
in the computation of the width.

\section*{Conclusions}

In theories with low string scale, there may exist longitudinal dimensions
(seen by the gauge interactions) in the TeV range. This possibility is also
favoured by supersymmetry breaking via the compactification, in four-dimensional
string theories. In the models discussed here quarks and
leptons are localized at particular points of the internal space. In contrast
gauge bosons have KK excitations.

In this paper we have studied the production, decays and signatures of these
KK-excitations at present (LEPII and Tevatron run-I) and future (NLC, 
Tevatron run-II and LHC) colliders. In some of them (LEP, NLC and Tevatron)
the energy is not high enough to produce on--shell KK modes and one should 
look at the deficit, induced by them, of cross-sections with respect to the
standard model ones. This indirect search would lead, in the case of no 
deviations, to lower bounds on the compactification scale that
can  be as high as $\sim$ 1.9 TeV for the case of LEPII and $\sim$ 8~TeV
and  13~TeV for NLC-500 and NLC-1000, respectively. 
On the other hand, the LHC can probe values of the order of 6.7 TeV (9 TeV) for one 
(two) large extra dimensions. These bounds (summarized in 
Table~\ref{thetable}) are very similar to those obtained for the quantum 
gravity scale~\cite{lowgrav} in the case of two large transverse dimensions.

\begin{table}[htb]
\begin{center}
\begin{tabular}{||c | c | c | c | c | c | c ||}\hline\hline 
Collider   &  LEPII  &  NLC-500  &  NLC-1000 &  TeV-I & 
TeV-II & LHC\\
\hline\hline
$R^{-1} ({\rm TeV}) \simlt $ &  1.9  & 8 & 13 & 0.9 & 1.2 & 6.7 \\ 
\hline\hline
\end{tabular}
\end{center}
\caption{\it Compactification scales that can be probed with 95\% confidence 
level at present and future colliders for one large extra dimension.}
\label{thetable}
\end{table}

At the LHC the energy can be high enough to produce the photon and $Z$ KK 
modes,
leading to a characteristic signal of resonances in the differential 
cross-section with respect to the invariant dilepton mass, which is the golden
signature for direct detection of these states. Although for large value of 
the compactification scale, one sees only the first KK-excitation resonance, 
this scenario can be differentiated from $Z'$ models by the observation of
excitations of the $W$ boson in $pp \rightarrow WX$ processes.

On top of these longitudinal dimensions with gauge interactions and their 
direct
production at high-energy colliders, there can exist transverse directions,
as large as (sub)millimeter, feeling only gravitational interactions. They
would indirectly contribute to the processes we have studied in this paper
(by exchange of towers of KK-excitations of gravitons and moduli fields) 
and provide independent signatures and
bounds on low string scales, which could be combined with ours. 

\vspace{2cm}
\begin{center}
{\large\bf Acknowledgements}
\end{center}
We acknowledge discussions with E. Accomando, F. Borzumati and particularly 
M.~Mangano and A.~Pomarol. KB thanks the IEM of CSIC (Spain) and the
CPhT of Ecole Polytechnique  (France) for hospitality. The work of IA and MQ
is supported in part by IN2P3-CICYT contract Pth 96-3. The work of KB is 
supported by a John Bell scholarship from the World Laboratory.

\end{document}